# The design and simulated performance of a coated nano-particle laser

**Joshua A. Gordon**

*College of Optical Sciences, University of Arizona, Tucson, AZ 85721*
*gordonj@email.arizona.edu*

**Richard W. Ziolkowski**

*Department of Electrical and Computer Engineering and College of Optical Sciences,*
*University of Arizona, Tucson, AZ 85721*
*ziolkowski@ece.arizona.edu*

**Abstract:** The optical properties of a concentric nanometer-sized spherical shell comprised of an (active) 3-level gain medium core and a surrounding plasmonic metal shell are investigated. Current research in optical metamaterials has demonstrated that including lossless plasmonic materials to achieve a negative permittivity in a nano-sized coated spherical particle can lead to novel optical properties such as resonant scattering as well as transparency or invisibility. However, in practice, plasmonic materials have high losses at optical frequencies. It is observed that with the introduction of active materials, the intrinsic absorption in the plasmonic shell can be overcome and new optical properties can be observed in the scattering and absorption cross-sections of these coated nano-sized spherical shell particles. In addition, a "super" resonance is observed with a magnitude that is $10^3$ greater than that for a tuned, resonant passive nano-sized coated spherical shell. This observation suggests the possibility of realizing a highly sub-wavelength laser with dimensions more than an order of magnitude below the traditional half-wavelength cavity length criteria. The operating characteristics of this coated nano-particle (CNP) laser are obtained numerically for a variety of configurations.

**OCIS codes:** (290.4020) Mie Theory; (140.3380) Laser Materials; (999.999) Nanophotonics; (999.999) Metamaterials.

# 1. Introduction

Optical metamaterials show the potential for realizing new, interesting and useful optical phenomena that can be designed to meet specific applications [1]. To realize some of the interesting properties of optical metamaterials, it is necessary for the electric permittivity (ENG metamaterials) or the magnetic permeability (MNG metamaterials) or even both of them (DNG metamaterials) to take on negative real values [2]. One of the conveniences of nature is that there are naturally occurring materials exhibiting negative permittivities at optical frequencies. On the other hand, one of the major obstacles in realizing metamaterials at optical frequencies is the lack of naturally occurring media exhibiting any magnetic response. It has been shown at microwave frequencies that a magnetic dipole moment can be induced and overall magnetic responses realized by imbedding in a non-magnetic host material, inclusions made from non-magnetic materials of appropriate designs (parallel nano-wires, split ring resonators, etc). It has also been demonstrated theoretically that by arranging nano-spheres in a ring configuration to create an "optical nano-circuit" [3], a magnetic dipole moment can be realized at optical frequencies via the displacement current induced in the ring by the incident optical field. Other interesting and desirable optical properties of nano-shells, which also require the use of negative permittivity and have been demonstrated, include resonant source and scattering configurations [4], as well as transparency and invisibility [5], [6-9]. Using the polarizability of the individual inclusions, one can realize effective electric permittivities and magnetic permeabilities that govern the electromagnetic response for waves interacting with the medium.

Recent work on resonant electrically-small concentric spherical shells [10-12] has demonstrated that these structures have highly tunable polarizabilities. At optical frequencies the sizes of these spherical structures are on the order of tens of nano-meters making them attractive candidates for use as inclusions in potential realizations of optical metamaterials. Current nano-fabrication capabilities have been used to successfully synthesize nano-shells, and many of their optical properties have verified experimentally [10]. To achieve the resonant tunability of spherical nano-shells at optical frequencies, one must incorporate plasmonic materials, such as metals, in the shells. Unfortunately, the polarizability of these structures is dominated by high losses at optical frequencies due to the absorption in these plasmonic materials. In an attempt to counter these intrinsic losses we have investigated the use of active media in multi-layered spherical plasmonic nano-shells.

There have been a number of recent studies, both theoretical and experimental, that have considered the influence of active media on nano-sized plasmonic particles, for instance, to overcome the large losses associated with metals at optical frequencies in scattering applications. The gain media considered have been generally dyes [13]-[16] and quantum dots [17]-[19]. The scattering configurations emphasized in these efforts have dealt with the partial or total immersion of the metallic nano-particles in the dyes or the placement of the metallic nano-particles in close proximity to the quantum dots. Most have dealt with silver as the metal; while others have emphasized gold. The use of rare-earth doped silica will be emphasized in this paper because this active medium will be surrounded by a metallic shell and, as a consequence, it may be the most compatible with existing coated nano-particle fabrication techniques.

We have used passive media models based on lossy dispersive materials to match the demonstrated properties of successfully synthesized passive plasmonic nano-shells. We have also developed active media models for several optical gain materials that have been successfully incorporated into these passive materials. We have then used these optical gain media to investigate their ability to overcome the losses associated with the spherical plasmonic nano-shells. We will demonstrate that a properly designed passive optical spherical core impregnated with a gain medium and coated with a concentric spherical plasmonic nano-shell will have a lasing state. The operating characteristics of this coated nano-particle (CNP) laser have been obtained numerically for a variety of configurations and will be reported here.

## 2. Electrically Small Resonant Scattering Structures

*2.1 Mie Theory*

The theory of plane wave scattering from an isotropic sphere was originally presented by Mie and extended to the more general case of concentric spherical shells by Aden [20] and others. For a linearly polarized plane wave incident on a concentrically layered spherical particle, the electric and magnetic fields in each region can be expanded into vector spherical harmonics. Enforcing the electromagnetic boundary conditions, i.e., the continuity of the tangential electric and magnetic fields at each interface, the scattered field coefficients are obtained.

From the scattered field and incident field the scattering cross-section and absorption cross-section are defined via Poynting's theorem. The scattering cross-section is defined as the total integrated power contained in the scattered field normalized by the irradiance of the incident field and the absorption cross-section is defined by the net flux through a surface surrounding the concentric shells normalized by the incident field irradiance, and is thus a measure of how much energy is absorbed by the concentric shell structure. The absorption and scattering cross-sections can be expressed via Poynting's theorem through the scattered and absorbed powers, which are given, respectively, by the following expressions:

$$P_{scat} = \text{Re}\left\{\frac{1}{2}\oiint_S \left[\vec{E}_s \times \vec{H}_s^*\right]\cdot\hat{n}\,dS\right\} \tag{1}$$

$$P_{abs} = -\text{Re}\left\{\frac{1}{2}\oiint_S \left[\vec{E}_{tot} \times \vec{H}_{tot}^*\right]\cdot\hat{n}\,dS\right\} \tag{2}$$

where $S$ is a sphere that surrounds the particle and $\hat{n}$ is the unit outward pointing normal to that surface. The incident, scattered, and total fields will be labeled by the subscripts *inc*, *scat*, and *tot*, respectively. The total scattering cross-section, absorption cross-section and extinction cross-section are thus defined from the ratio of the scattered or absorbed power to the incident irradiance $I_{inc,}$ and can be expressed in terms of the scattered field coefficients as:

$$\sigma_{scat} = \frac{P_{scat}}{I_{inc}} = \frac{2\pi}{\beta_o^2}\sum_n^\infty (2n+1)\left(|a_n|^2 + |b_n|^2\right) \tag{3}$$

$$\sigma_{abs} = \frac{P_{abs}}{I_{inc}} = -\frac{2\pi}{\beta_o^2}\sum_n^\infty (2n+1)\left(\text{Re}\{a_n\} + |a_n|^2 + \text{Re}\{b_n\} + |b_n|^2\right) \tag{4}$$

$$\sigma_{ext} = \sigma_{scat} + \sigma_{abs} \tag{5}$$

where $\beta_o = 2\pi/\lambda$ when the particle is embedded in free space, $\lambda$ being the wavelength of the source. The scattering and absorption efficiencies are then defined as the ratio of the corresponding cross-section to the geometric cross-section of the particle.

$$Q_{scat} = \frac{\sigma_{scat}}{\pi r_2^2} \tag{6}$$

$$Q_{abs} = \frac{\sigma_{abs}}{\pi r_2^2} \tag{7}$$

$$Q_{ext} = Q_{scat} + Q_{abs} \tag{8}$$

where $r_1$ and $r_2$ are, respectively, the inner and outer radii of the shell respectively.

The scattered field coefficients can be determined by solving a system of equations derived from matching the tangential components of the electric and magnetic fields at the boundaries of the inner and outer surface of the spherical shell as in [20]. The resulting matrix equation takes the form:

$$[M] \cdot [C] = \begin{bmatrix} M_{TE} & 0 \\ 0 & M_{TM} \end{bmatrix} \cdot \begin{bmatrix} A_{TE} \\ B_{TM} \end{bmatrix} = \begin{bmatrix} f_{TE} \\ f_{TM} \end{bmatrix} = [F] \tag{9}$$

where $[M]$ is the scattering matrix, which consists of combinations of the expansion functions and their derivatives evaluated at the boundaries; $[C]$ is a vector containing the coefficients of both the TE and TM field components in each region; and $[F]$ is the forcing vector defined by the incident plane wave at the outer boundary of the particle. As indicated, the scattering matrix can be expressed as a block diagonal matrix with sub-matrices which individually describe the TE and TM fields. Consequently, the TE and TM scattered field coefficients can be obtained independently. Applying Cramer's rule, the scattered field coefficients can be expressed as the ratio of determinants:

$$a_n^{TE} = \frac{\det[\tilde{M}_{TE}(f_{TE})]}{\det[M_{TE}]}$$

$$b_n^{TM} = \frac{\det[\tilde{M}_{TM}(f_{TM})]}{\det[M_{TM}]}$$

(10)

where $\tilde{M}_{TE}(f_{TE})$ and $\tilde{M}_{TM}(f_{TM})$ are the TE and TM sub-matricies with the first column replaced by the sub-vectors $f_{TE}$ and $f_{TM}$, respectively. The coefficients thus depend explicitly on the material functions $\varepsilon$ and $\mu$ in each region; the inner and outer radii of the CNP shell: $r_1$ and $r_2$; and the wavelength. It is apparent that when the determinant in the denominator approaches a minimum, a resonance in the scattering parameters of these CNPs and, as a result, for example, their total cross-section can occur. It has been shown [10, 11, 21] that the resonance condition depends on the *ratio* of the core radius to the total particle radius as well as on the properties of the core, shell and surrounding medium. Aside from their existence, a very attractive characteristic of these electrically-small resonances is their explicit dependence on the shell radii. This property allows for the tunability of their frequencies by changing the geometry. Consequently, these tunable electrically-small resonances are of interest when considering these CNPs for applications, as well as for inclusions in optical metamaterials. The design of the CNPs described below was thus accomplished for a desired value of the resonance wavelength by determining the radii ratio for given core and shell materials. To illustrate the design and tunability of a CNP, Log plots (base10) of the absolute value of $a_n^{TE}$, $b_n^{TM}$, $\det[M_{TE}]$ and $\det[M_{TM}]$ are shown in Fig. 1 as functions of the ratio of $r_1$ and $r_2$ at $\lambda = 510nm$ when $r_2$ is fixed at $r_2 = 30nm$ for a silver nano-shell surrounding a silica nano-core, which has $\varepsilon = 2.05\varepsilon_0$. In the design of the CNPs, the dependence of the resonance on losses was also investigated. The determinant of the denominator for the shell material being modeled with both a lossy and lossless bulk silver dielectric function are compared in the bottom right plot in Fig. 1(d). From this result one notes that there is a negligible, if any affect on the resonant geometry when losses are introduced into a CNP design.

At $\lambda = 510nm$, it is seen that the denominator in the TM coefficients attains a minimum at the radii ratio, Rr = 0.8, which corresponds, respectively, to a core radius and outer radius of 24nm and 30nm. This minimum coincides with the maximum in the TM scattering coefficients $b_n^{TM}$. We note that this minimum is caused by the fact that the imaginary part of the determinant goes to zero at this radii ratio while there is no dramatic change in the real part. Notice that the TE scattering coefficients, $a_n^{TE}$, do not exhibit a resonance, but they do have a maximally reduced resonance near the radii ratio of 0.97. Similarly, $b_n^{TM}$ has a maximally reduced resonance near the radii ratio 0.95. Such non-scattering states have recently gained attention for use in metamaterials for achieving invisibility [5]. The wavelength tunability of the electrically small resonances associated with varying the radii ratio of the CNPs is demonstrated in Fig. 2, with the silver-silica CNP. Due to the losses inherent in silver, as will be discussed below, the designs favorable for active CNPs using silver shells fall near $\lambda = 510nm$. Thus we have selected the radii ratio of Rr=0.8 for all of the silver-based CNP simulations reported here.

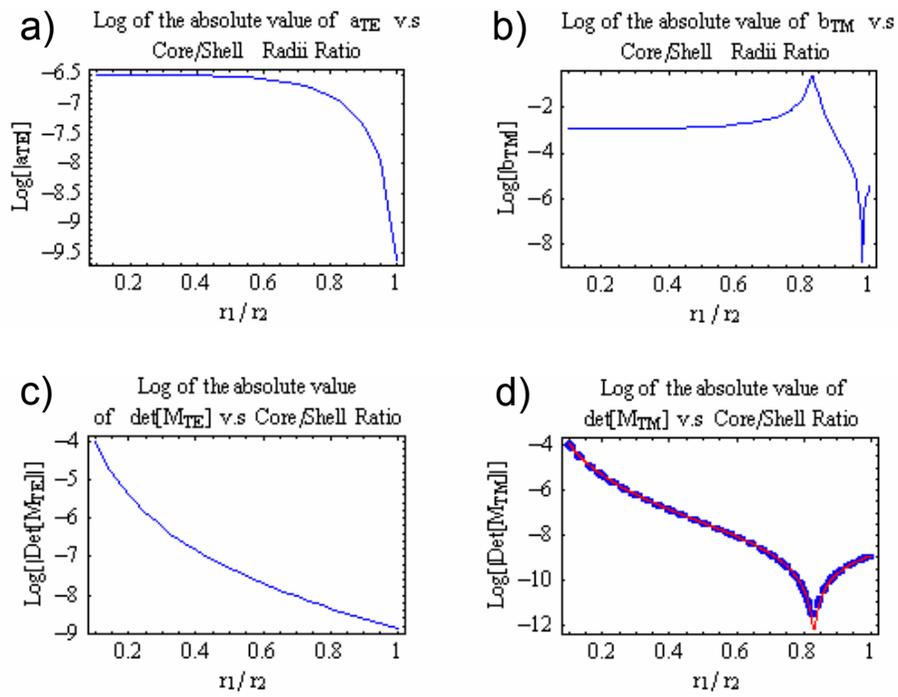

Fig. 1. Dependence of the terms in the scattering coefficients as a function of the radii ratio for a Ag-SiO$_2$ CNP, a) TE coefficient, b) TM coefficient, c) TE coefficient denominator, and d) TM coefficient denominator. In d) both lossless (red, solid curve) and lossy (blue, dashed curve) bulk Ag results are given.

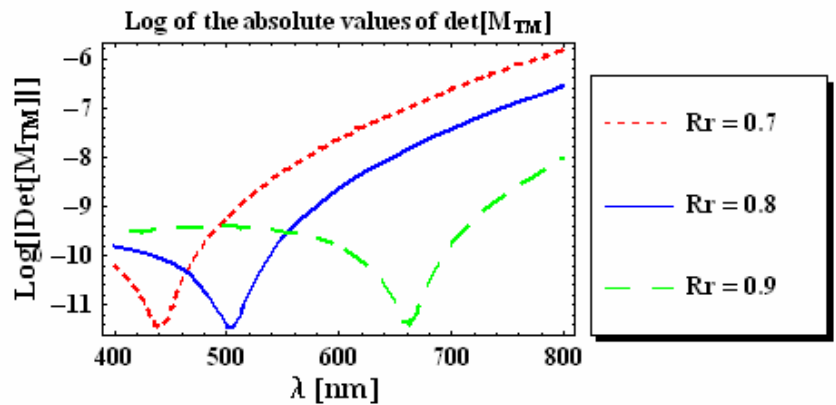

Fig. 2. Wavelength tunability of the TM resonance of the Ag-SiO$_2$ CNP by varying the radii ratio.

*2.2 Resonant Electrically Small Spheres*

In recent years there have been several investigations on the effects of multi-layered metamaterial spheres and their use for enhancing electromagnetic phenomena. At radio and microwave frequencies, it has been shown that large enhancements are attainable in the radiated power from a dipole antenna located arbitrarily near to or centrally within concentric spherical metamaterial shells [12], [22-24]. In addition to enhancing the radiated power it has also been found that there are situations where there is a drastic reduction in the radiated power from the dipole antenna, i.e., there exist not only enhanced radiating states but also non-radiating states [21, 25].

Analogous to the radiated power enhancements obtained with concentric spherical metamaterial shells, it has also been shown that there are reciprocal enhanced scattering states [23]. These resonant scattering states have been recognized previously [10, 21]. Analogous to the non-radiating states associated with the radiating dipole antenna in the presence of a metamaterial shell, extremely low scattering states also exist. In fact a great deal of recent attention has been given very recently to the so-called "transparency" or "invisibility" effect [5, 6-9, 11].

*2.3 Optical Excitations*

Recent work has investigated the use of optically tunable resonant passive nano-spheres for biomedical and optical applications. Projected uses for these nano-shells range from exploiting their tunability for contrast agents in early cancer detection [26, 27] and for drug delivery, to creating near infrared, highly absorbing particles for use in a photo thermal ablation therapy for cancer treatment [27]. Optically tunable plasmonic nano-shells have been successfully synthesized with spherical dielectric cores surrounded by thin metallic coatings for a number of years. Dielectric-core gold nano-shells have been successfully synthesized, and their optical tunability has been verified experimentally [10]. In these particles gold nano-shells with thicknesses as thin as 2nm surround a gold sulfide $AuS_2$ core. Tunability from 550nm to 950nm was demonstrated using particle radii ranging from 4nm to 17nm, respectively. Tunable gold nano-shells made with a silica core have also been successfully synthesized and their optical properties experimentally verified [27]. Tunability was demonstrated for larger spheres with a 60nm core size and shell thicknesses ranging from 5nm to 20nm to cover the wavelength range from 750nm to 1000nm in which the resonance was observed to shift to longer wavelengths as the core to shell radius ratio increased. In addition to gold-silica nano-shells, silver coated silica nano-shells with core radii of 40-250nm and shell thicknesses of 10- 30nm have also been successfully synthesized and optical properties experimentally verified [28]**.** It was demonstrated that these silver-silica nano-shells exhibit tunable plasmon resonances at shorter wavelengths with a 10% larger enhancement than for gold-silica nano-shells and, therefore, can be used to cover a wider portion of the optical spectrum.

*2.4 Nano-Scatterers*

In this section the optical characteristics of nano-shells will be discussed in more detail. Optical nano-shells possess tunable resonances where the enhancements in the extinction cross-section are attributed to the optical field coupling to plasmon modes of the metal shell. In the cases of lossless materials the extinction cross-section is dominated by light scattered by the particle and is equal to the scattering cross-section. Material absorption must be present for a finite absorption cross-section and only then will it contribute to the extinction cross-section of the particle. When material absorption is present enhancements in the absorption cross-section of the particle can occur as well. The intrinsic properties dominate the extinction in the quasi-static regime, $\beta_0 a \ll 1$, where the scattering cross-section and absorption cross-section are dominated by the dipole terms. Differences in the contributions to the extinction cross-section from scattering and absorption result from extrinsic and intrinsic optical properties of the nano-shells. Intrinsic effects dominate in the quasi-static regime and extrinsic effects become non-negligible for nano-shells with dimensions larger than the quasi-static limit. Larger spheres that incorporate lossy materials tend to be dominated by scattering and less by absorption when compared to smaller nano-shells of similar structure. Nano-shells which fall into the quasi-static regime are mostly dominated by absorption due to predominant coupling to the lowest order dipole plasmon mode. For larger spheres, phase retardation effects are more significant and therefore optical fields can couple to higher order plasmon resonances, which correspond to higher order multipole fields. This increase in coupling between the optical field and plasmon modes of the nano-shell results in extinction cross-sections dominated by scattering rather than by absorption for larger particles [29, 30]. The extinction cross-section spectrum in these two size regimes is different as well. For smaller spheres only one plasmonic resonance is dominant. On the other hand, for a larger particle not in the quasi-static regime, i.e., where $\beta_0 a \geq 1$, higher order plasmon resonances corresponding to higher order multipoles can be excited in the scattered fields. Consequently, multiple resonances can occur. Due to the increases in the scattering and phase retardation effects, the widths of the resonances are wider in larger particles than they are for smaller spheres [29, 30]. Such differences in absorption and scattering are important when considering lossy plasmonic nano-shells for optical regime applications.

In our investigations of active nano-shells we felt it was important to consider optical materials that not only have properties suitable for realizing resonant nano-shells, but that also have been shown to be realistically synthesized into nano-shells. As mentioned above nano-shells have been successfully created using combinations of gold and silver as materials for the metal shells and of gold sulfide and silica as the dielectric core material. For this reason we have used silver, gold and silica in our models. In modeling these nano-shells both the extrinsic and intrinsic optical properties of these nano-sized particles must be considered.

## 3. Optical Material Properties

*3. 1 Passive media*

Due to the nano-scale dimensions of the particles under investigation, accurate modeling of the optical properties of the nano-shells requires that one takes into account the size dependence of the materials used in making these structures. The size dependence of the optical properties of nano-scale particles can be classified as either extrinsic, i.e., if the size effects arise predominantly from electro-dynamic effects, or intrinsic, i.e., if there is an actual change in the optical response of the materials that comprise the particle. In our layered nano-particles the gold and silver plasmonic shells exhibit significant intrinsic size dependencies. We have used empirically determined bulk values for the permittivity of gold and silver at optical wavelengths between 200nm to 1800nm [31, 32]. Following the approach in [30], the permittivity is decomposed into a size dependent Drude response and an interband transition response as follows:

$$\varepsilon(\omega, R) = \varepsilon_{Drude}(\omega, R) + \chi_{IntBand}(\omega) \quad (11)$$

where the term $R$ is the thickness of the metal shell and the Drude permittivity is given by the expression,

$$\varepsilon_{Drude}(\omega, R) = 1 - \frac{\omega_p^2}{\Gamma(R)^2 + \omega^2} + i\frac{\Gamma(R)^2 \omega_p^2}{\omega(\Gamma(R)^2 + \omega^2)} \quad (12)$$

where $\omega_p$ amd $\Gamma$ are, respectively, the plasma and collision frequencies.

The size dependence is treated as an effect which arises when the size of the material approaches the bulk mean free path length of the conduction electrons in the material. It is treated as an alteration in the mean free path which is then incorporated into the Drude model as a size dependent damping frequency. In particular, the damping frequency is assumed to take the form,

$$\Gamma(R) = \Gamma_\infty + \frac{A\,V_F}{R} \quad (13)$$

where A is a constant term assumed to be approximately unity, i.e., $A \sim 1$. The term $V_F$ is the Fermi velocity. The Drude parameters and the Fermi velocity values used in our simulations for silver and gold are given in Table 1,

**Table 1. Gold and Silver material model constants**

|  | $m^*/m$ Kg | N $10^{28} m^{-3}$ | $\omega_p$ $10^{16} s^{-1}$ | $\Gamma_\infty$ $10^{13} s^{-1}$ | $V_F$ $10^6 m/s$ | A |
|---|---|---|---|---|---|---|
| Gold | 0.99 | 5.90 | 1.39863 | 3.30952 | 1.39 | 1 |
| Silver | 0.96 | 5.85 | 1.39269 | 2.67308 | 1.39 | 1 |

In addition to increasing the real part of the permittivity, the most significant size dependent effect of consequence for considering resonant nano-shells is the large increase in optical loss as the material size decreases to dimensions on the order of tens of nanometers. This is shown in Figs. 3 and 4, for both gold and silver. The blue line in each figure depicts the permittivity values for the bulk metal, and the black lines indicate increasing material dimensions ranging from 2-100nm. The figures indicate that as the dimension decreases, the magnitude of the real and imaginary parts increase. Due to the reduction of the mean free path of the Drude electrons in the thin nanometer thick shells, the collision frequency increases. Therefore, more of the kinetic energy is dissipated as heat, which results in an increase in the optical loss. This increased optical loss must be taken into account when one chooses an appropriate medium with sufficient gain to compensate for the losses of the size dependent plasmonic shells.

The size dependence of the plasmonic metal shells alters the resonance characteristics of a passive nano-shell. The most notable effect is a broadening of the resonance and a corresponding reduction in its strength. When a bulk dielectric function is used for the shell model, the position of the resonance in not significantly affected by the size dependent properties. Comparing the bulk and size dependent models of the shells, one finds that the resonance position of the complete model is shifted significantly when the interband transition contributions are neglected and only the Drude component of permittivity is considered. This property is shown in Figs. 5 and 6, where the absorption and scattering efficiencies for gold coated and silver coated silica passive nano-shells are shown. The results for the Drude, bulk, and size dependent permittivity models are presented. Because of the large amplitude differences in these cases, these efficiency values were also normalized to one. These normalized values are also presented in these figures; they clearly show the locations of the resonances predicted with these models. The size dependent effects are seen in both the scattering and absorption cross-sections. For the Au-SiO$_2$ case, the shell thickness is 2nm with $r_1 = 22.5$ nm and $r_2 = 24.5$ nm, while the shell thickness is 6nm for the Ag-SiO$_2$ case with $r_1 = 24.0$ nm and $r_2 = 30.0$ nm. These results strongly emphasize that using only a size-independent Drude model to simulate the material properties of a CNP neglects important physical effects which significantly impact its scattering and absorption efficiencies.

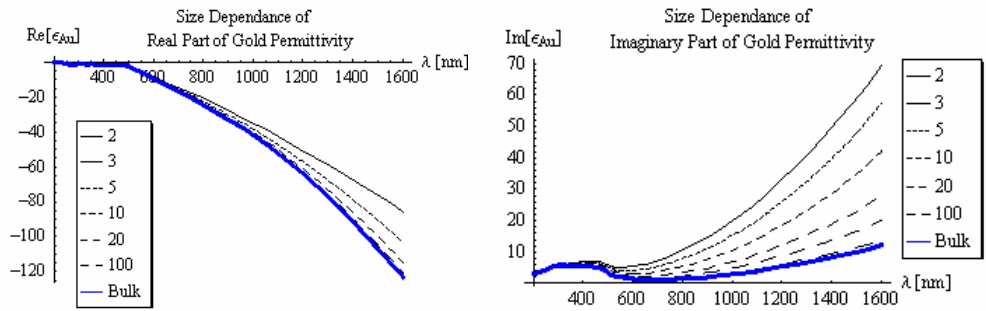

Fig. 3. Size dependence of the permittivity of gold.

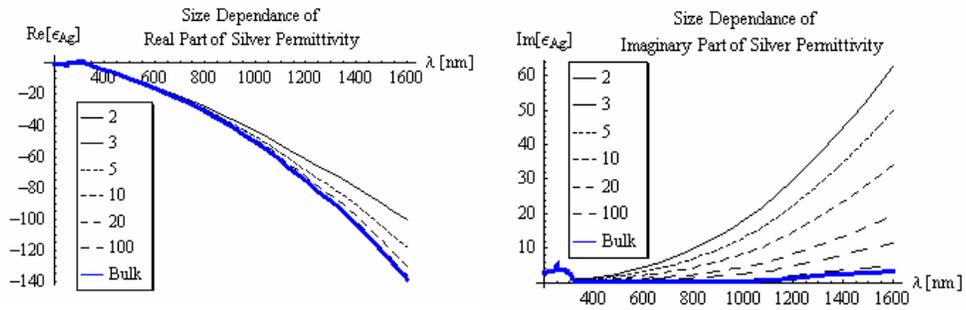

Fig. 4. Size dependence of the permittivity of silver.

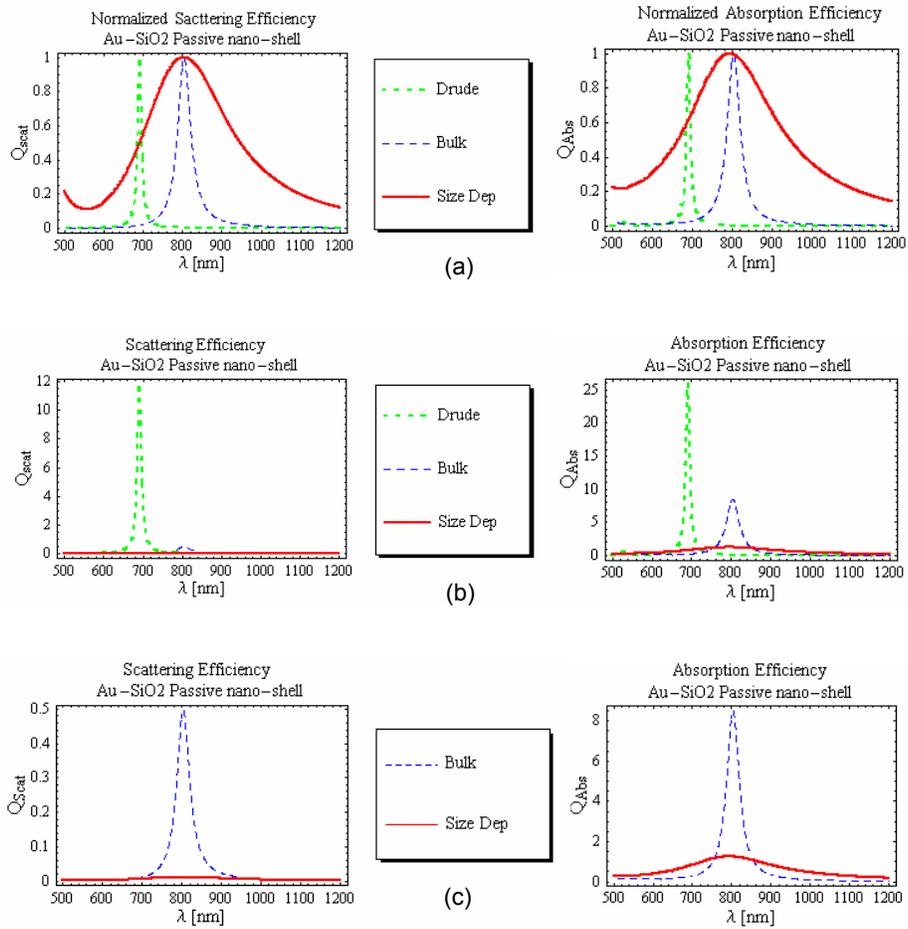

Fig. 5. The effects that different Au models have on the efficiencies for the Au-SiO$_2$ passive CNPs are compared. Drude, bulk, and size dependent models of the Au are shown. a) Comparison of the normalized efficiencies to show the location of the resonances, b) Comparison of the unnormalized efficiencies to show the dominance of the Drude results, and c) Comparison of the unnormalized bulk and size dependent efficiencies to show that the size dependent, i.e., the most physical nano-scale model, results produce the lowest level, largest bandwidth resonance.

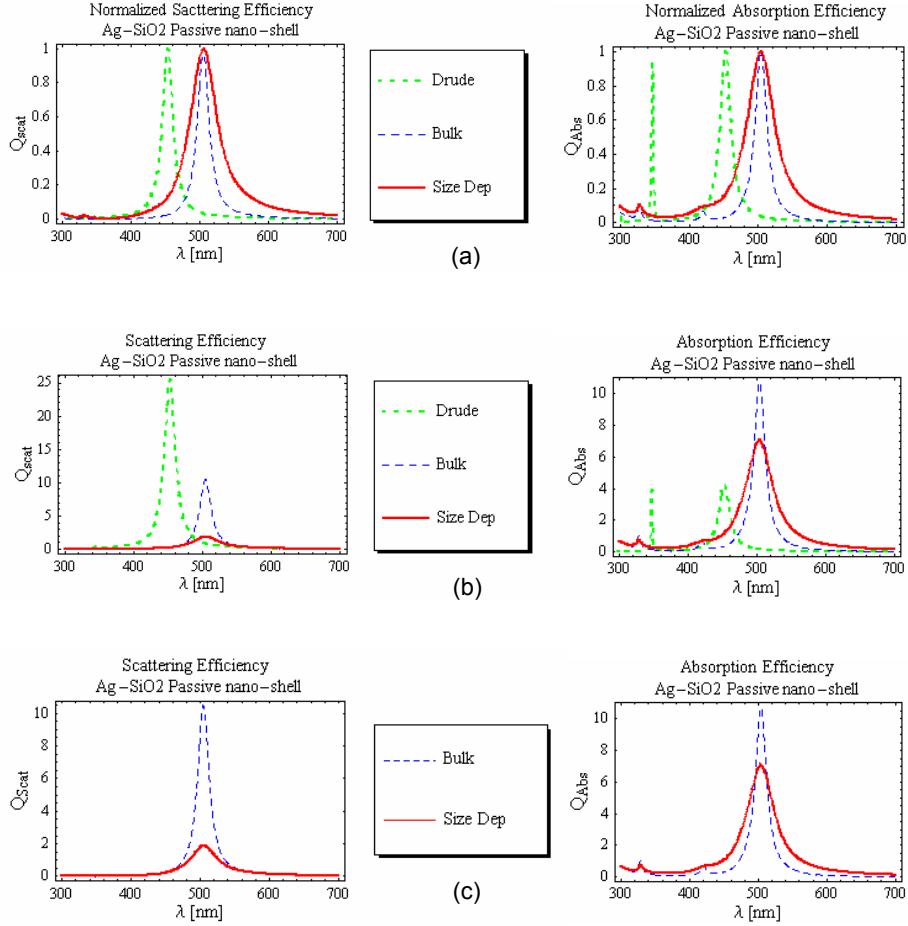

Fig. 6. The effects that different Ag models have on the efficiencies for the Ag-SiO$_2$ passive CNPs are compared. Drude, bulk, and size dependent models of the Ag are shown. a) Comparison of the normalized efficiencies to show the location of the resonances, b) Comparison of the unnormalized efficiencies to show the dominance of the Drude results, and c) Comparison of the unnormalized bulk and size dependent efficiencies to show that the size dependent, i.e., the most physical nano-scale model, results produce the lowest level, largest bandwidth resonance.

In modeling active nano-shells we have considered both canonical gain models and models that resemble those which have been successfully incorporated into materials used in the synthesis of passive optical nano-shells. For our investigations of the core permittivity parameter space for the theoretical gain medium, we have used a general permittivity model described in terms of the real part of the refractive index, *n*, and the imaginary part of the refractive index, *k*, which represents the optical loss/gain constant. The permittivity in this model is thus defined as:

$$\varepsilon = n^2 - k^2 + i2kn \tag{14}$$

For the applications of this model below *n* is maintained at the silica, SiO$_2$, index value of 1.431. The values of *k* are varied over a range of values that represent the expected loss or gain in the core of the nano-shell. The results obtained by using such a model will be presented below where it will be shown that there exist resonance states of enormously enhanced radiated power for certain coordinates in the refractive index and optical gain parameter space.

*3.2 Active media*

Many rare-earth ions are known that can be used as dopants in a dielectric host material to achieve optical gain including, Pr$^{3+}$, Ho$^{3+}$, Er$^{3+}$, Nd$^{3+}$, Tm$^{3+}$. These ions can provide gain over different wavelength regions spanning from the visible to the near infrared [33, 34]. In telecommunication technologies, for instance, doping silica with the rare-earth ion Erbium (Er$^{3+}$) has been proven as an effective means of realizing gain in a silica host material for some time. Many rare-earth ions in the presence of a dielectric host can be modeled as a three-level Stark-split atomic system. Following [35] we have introduced gain by considering a susceptibility model suitable for representing such a three level system. Due to the complex permittivity values of gold, the passive CNPs constructed with gold shells are restricted to longer resonance wavelengths than those constructed with silver shells. At these longer wavelengths the optical losses for gold are larger than for silver. As will be demonstrated below with the canonical gain model, the required gain needed to overcome the losses in the structures consisting of silver shells is considerably less than for gold, i.e., $\left|k_{Ag}\right| < \left|k_{Au}\right|$. As a result, we have focused our study on active silver CNPs to investigate the active doped glass parameters needed to realize the gain required in order to overcome the losses in active CNP structures. Moreover, to achieve electrically small active CNPs in the visible, the susceptibility of the rare-earth gain model considered below will be driven at 510nm, which falls within the region where the optical loss constant of silver is lowest and within the region where gain lines of several rare earth ions are available.

The total complex optical susceptibility can expressed as a sum of the background susceptibility due to the host material and additional contributions due to the rare-earth ions via the total material polarization, $P_T$ ,

$$\begin{aligned}
P_H &= \varepsilon_0(1+\chi_H)E_{signal} \equiv \varepsilon_0 n^2 E_{signal} \\
P_{ion} &= \varepsilon_0 \chi_{ion} E_{signal} \\
P_T &= P_H + P_{ion}
\end{aligned} \quad (15)$$

where $E_{signal}$ is the field interacting with the gain medium and the rare-earth ion and total susceptibilities are

$$\chi_{ion}(\lambda) = \chi'_{ion}(\lambda) - i\chi''_{ion}(\lambda) \quad (16)$$

$$\chi_T(\lambda) = \chi_H(\lambda) + \chi_{ion}(\lambda) \quad (17)$$

When the time scale of the atomic excitation is long compared to the time of the thermally assisted transitions for each of the Stark-split levels, the rare-earth ion-based contribution to the susceptibility can be expressed as a superposition of the atomic susceptibilities associated with each level in the stark manifold [35, 36]. The result is an expression for the susceptibility in terms of the absorption and emission cross-sections for the rare-earth ions. The real and imaginary parts of the susceptibility are related to the absorption and emission cross-sections via the expressions:

$$\chi'_{ion}(\lambda) = n\frac{\lambda}{2\pi}\left[\sigma'_e(\lambda)\bar{N}_2 - \sigma'_a(\lambda)\bar{N}_1\right] \quad (18)$$

$$\chi''_{ion}(\lambda) = -n\frac{\lambda}{2\pi}\left[\sigma''_e(\lambda)\bar{N}_2 - \sigma''_a(\lambda)\bar{N}_1\right] \quad (19)$$

where

$$\bar{N}_1 = \frac{N}{1+P} \quad , \quad \bar{N}_2 = \frac{NP}{1+P} \quad (20)$$

and $\sigma''_e(\lambda)$ and $\sigma''_a(\lambda)$ are, respectively, the emission and absorption cross-sections of the rare earth ion, which can be obtained from absorption and fluorescence spectra; $P$ is the ratio of the pump power to the threshold power of the rare-earth ion; $n$ is the real part of the host medium refractive index; and $N$ is the concentration of the rare-earth ions. The cross-sections, $\sigma'_e(\lambda)$ and $\sigma'_a(\lambda)$ used in determining the real part of the susceptibility, are determined through a Hilbert transform via the Kramers-Krönig relationships as,

$$\sigma'_e(\lambda) = \frac{1}{\pi}P.V.\int_{-\infty}^{\infty}\frac{\sigma''_e(\omega)}{\omega' - \omega}d\omega'$$

$$\sigma'_a(\lambda) = \frac{1}{\pi}P.V.\int_{-\infty}^{\infty}\frac{\sigma''_a(\lambda)}{\omega' - \omega}d\omega' \quad (21)$$

In the following calculations, the normalized emission and absorption cross-section spectra obtained from [36], which are representative of rare-earth-doped silica, were used. As can be seen from Eqs. (16)-(20), the magnitude of the total susceptibility at a given pump power ratio, $P$, is highly dependent on the product of the doping ion concentration, $N$, and the absorption, $\sigma''_a(\lambda)$, and emission, $\sigma''_e(\lambda)$, cross-sections. For example the optical susceptibility used for the active silver CNP in the following simulations was obtained with

$N\sigma_{em} = 5.8\times 10^4 cm^{-1}$ and $N\sigma_{abs} = 5.3\times 10^4 cm^{-1}$. Susceptibility values for different pumping powers are shown in Fig. 7. It is observed that as the pump power ratio varies, both the real and imaginary parts of the susceptibility are affected.

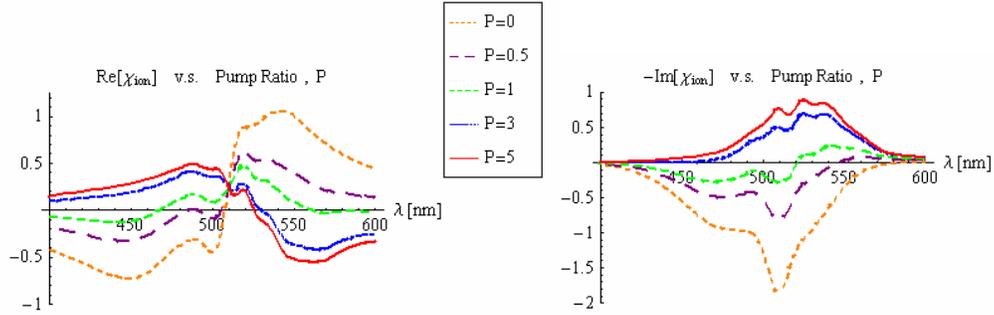

Fig. 7. The real and imaginary parts of the rare-earth ion doped silica susceptibility, with the parameters $N\sigma_{em} = 5.8\times 10^4 cm^{-1}$ and $N\sigma_{abs} = 5.3\times 10^4 cm^{-1}$ as the pump power ratio, $P$, is varied.

**4. Coated Nano-Particles**

*4.1 Passive CNPs*

We have investigated the optical properties of nano-shells comprised of gold and silver shells and active core materials modeled as rare-earth doped $SiO_2$. In the following sections the optical properties of these active CNPs will be presented and compared to the passive case of a pure silica core. For passive CNPs with lossy plasmonic shells, the extinction cross-section is dominated by absorption. This is true even for very thin shells, when the shell thickness is much less than the total particle radius. Examples of this are shown in Fig. 8, for the cases of an Au-$SiO_2$ CNP with a shell thickness of 2nm, and for an Ag-$SiO_2$ CNP with a shell thickness of 6nm. For these and all of the following results, the size dependence of the metal shells and its impact on the permittivity was taken into account. We will show that active materials can compensate for these losses and even overcome them so that the resulting extinction cross-section is dominated entirely by radiated power. In addition, we will show that in this regime where the extinction cross-section is dominated by the radiation, optical phenomena exist which we have interpreted as the onset of lasing in an active CNP.

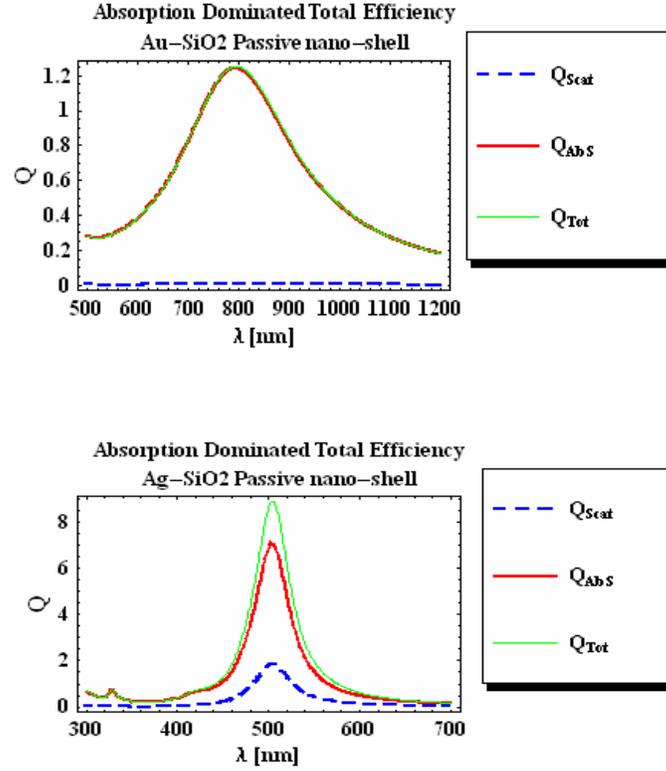

Fig. 8. Comparisons of the contributions of the scattering and absorption efficiencies demonstrate that the total efficiency in passive Au-SiO$_2$ and Ag-SiO$_2$ CNPs is dominated by the absorption.

*4.2 Active CNPs*

To model the active CNPs with the canonical active permittivity given by Eq. (14), the index of silica was taken to be n=1.431 and the optical gain constant, *k*, was varied in the interval: $-1 \leq k \leq 0$. These parameters provided coverage over the domain of the resonant passive CNPs and allowed us to explore the incremental changes in the scattering properties as the gain was varied. The scattering and absorption cross-sections were calculated for both Au-SiO$_2$ and Ag-SiO$_2$ CNPs. The Au CNP had a core radius $r_1 = 22.5$ nm and a 2nm thick shell, and the Ag CNP had a core radius $r_1 = 24.0$ nm and a 6nm shell. For the canonical gain model, the resulting scattering and absorption cross-sections for the Au-SiO$_2$ CNP are plotted in Figs. 9 and 10. Similar behavior was observed in the Ag-SiO$_2$ CNP case.

As defined by Eqs. (2) and (4), the absorption cross-section $\sigma_{abs}$ is a measure of the net outward power flux scattered from the CNP, where a positive $\sigma_{abs}$ indicates power lost due to absorption within the CNP. Therefore, a negative absorption cross-section is interpreted as a net power leaving the CNP, i.e., the CNP has become a nano-radiator with a projected radiant exitance equal to the incident irradiance scaled by the extinction efficiency of the CNP. At the point where $\sigma_{abs}$ becomes zero, the losses associated with the passive CNP have been compensated by the gain. As $\sigma_{abs}$ becomes more negative, the total amount of light leaving the region of the CNP increases, which means that the scattered radiation is now accompanied by power being radiated by the active CNP. In the following, data will be presented as the scattering and absorption efficiencies ($Q_{abs}$ and $Q_{scat}$), as defined by Eqs. (6), (7), and (8).

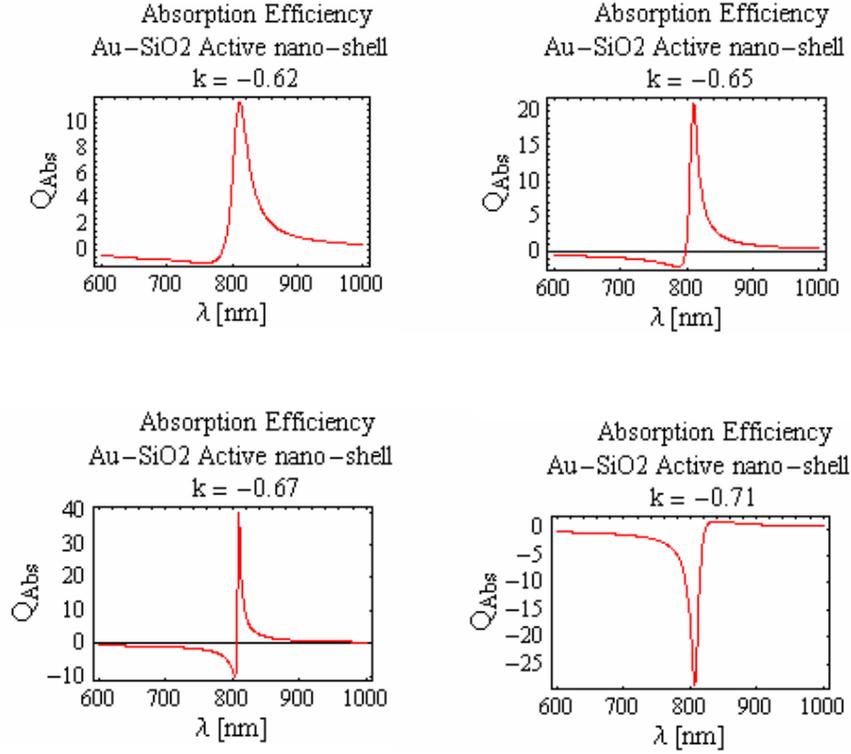

Fig. 9. Absorption efficiencies for the Au-SiO$_2$ CNPs for several values of the loss/gain parameter $k$.

As $Q_{abs}$ decreases, $Q_{scat}$ increases. However, $Q_{scat}$ does not increase monotonically with a decrease in $Q_{abs}$. One finds that $Q_{scat}$ begins to grow near the point where the value of $Q_{abs}$ goes through zero. In addition to the strength of $Q_{abs}$ and $Q_{scat}$ being affected by gain present in the core, the widths of the scattering and absorption resonances change as well. As the gain increases to the point where $Q_{abs}$ becomes negative, the width of $Q_{scat}$ also narrows. As the gain continues to increase, $Q_{scat}$ broadens out again. The narrowest scattering resonances coincide with the largest $Q_{scat}$ values. This narrowing and then broadening of $Q_{scat}$ follows the non-monotonic nature of the maximum values of $Q_{scat}$ as the gain is increased past the point where the losses associated with the passive CNP are overcome. It is also found that by varying the gain in the core, the scattering and absorption efficiencies of the CNP can be adjusted. Consequently, it can be envisioned that by adjusting the pump level applied to the gain medium (for any given pumping scheme), the absorption of an active CNP and, hence, a metamaterial comprised of such active CNPs, could also be adjusted dynamically.

For our Au-Silica CNP calculations, it was found that as the gain is increased, the absorption decreases until the value of $k$ is in the neighborhood of -0.67. In that neighborhood $Q_{abs}$ becomes negative while $Q_{scat}$ increases dramatically.

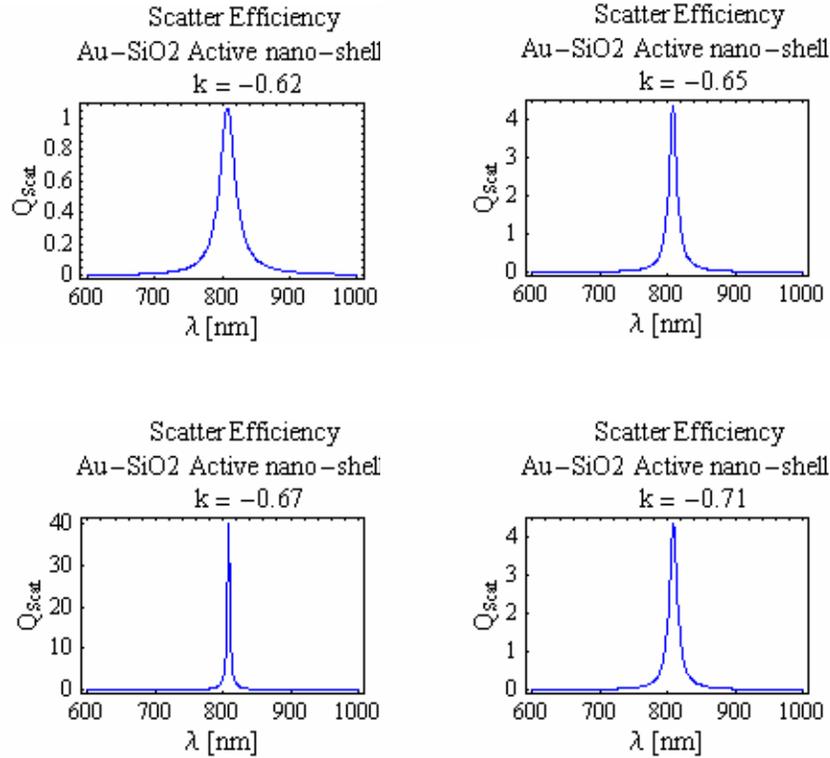

Fig. 10. Scattering efficiency for the Au-SiO$_2$ CNPs for several values of the loss/gain parameter $k$.

Further investigation of the $k$ values near to where $Q_{scat}$ is maximized shows that the resonance of the CNP can become extremely large, i.e., the peak value is several orders of magnitude larger than for any other $k$ values. In fact both $Q_{abs}$ and $Q_{scat}$ attain extrema at those critical values of $k$. Particularly intriguing is the fact that $Q_{abs}$ in this region is negative, indicating that a large amount of power is radiated from the CNP. Along with the large resonances, the widths of both $Q_{scat}$ and $Q_{abs}$ become extremely narrow, going from several hundred nanometers down to only a few tens of nanometers. These features are illustrated in Fig. 11.

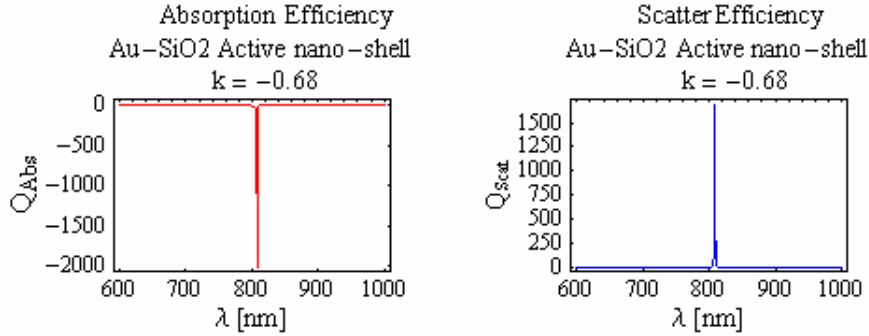

Fig. 11. The absorption and scattering efficiencies for an Au-SiO$_2$ CNP with $k$ slightly beyond critical. The narrowing of the width of these efficiencies is immediately apparent.

At $k = -0.68$, both $Q_{scat}$ and $Q_{abs}$ attain magnitudes on the order of $10^3$, with a full-width-at-half-maximum (FWHM) of about 10nm. As indicated by the definitions (5) and (8), a measure of the peak net power leaving the vicinity of the CNP is calculated as the difference between $Q_{scat}$ and $Q_{abs}$, i.e., the total efficiency. The Log (base 10) of the absolute value of this total efficiency is plotted as a function of the gain parameter, k, for $-1 < k < 0$, in Fig. 12a. It is clear from this result that the active CNP resonance passes through a large enhancement over the region $-0.8 < k < -0.6$, with a large positive net power being radiated away from the CNP. The sharp inflection near the coordinate (-0.7, 1) is due to the total efficiency crossing the zero point, i.e., the point for which the losses have just been overcome and $Q_{abs}$ changes from positive to negative values, and our choice to plot the Log of the absolute value. The values to the right of this point for $k > -0.7$ represent a negative total efficiency and, therefore, a net absorption by the CNP. Similar results for the case of the active Ag-SiO$_2$ CNP were found and are shown in Fig. 12b. It is observed that the gain parameter $k$ required to overcome the losses is less in the Ag case. In particular, the inflection point near (-0.15, 0) corresponds to the $k$ value below which the losses are overcome and the absorption efficiency becomes negative. The critical value of $k$ needed to achieve a super resonance with negative absorption efficiency for the Au case is nearly three times that of the Ag case, i.e., for the Ag case the super resonance occurs with $k=-0.25$ and for the Au case it occurs with $k = -0.68$. Thus, the optical gain coefficient $\alpha = 2\pi k / \lambda$ is $-5.28 \times 10^4 cm^{-1}$ for Au at 809 nm, and is $-3.08 \times 10^4 cm^{-1}$ for Ag at 510 nm.

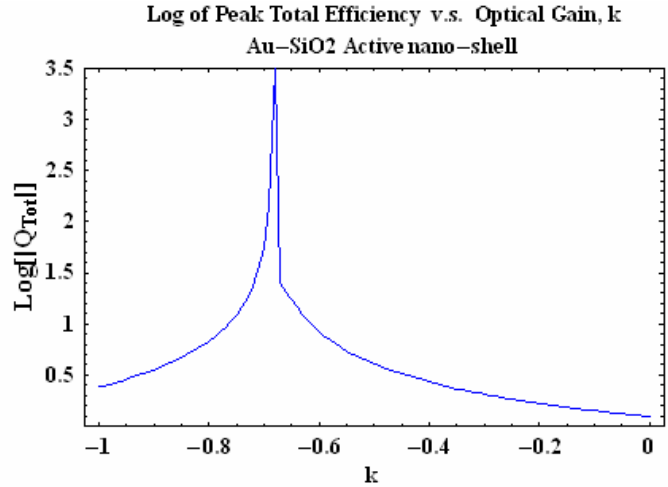

(a)

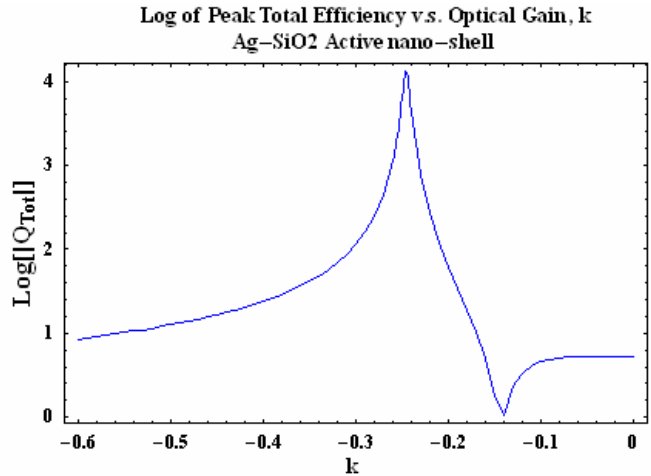

(b)

Fig. 12. Comparison of the optical gain constant values of (a) Au, and (b) Ag, that are required to overcome the passive CNP losses and to achieve the super-resonant state. Shown is the Log (base 10) of the absolute value of the total efficiency as a function of the optical gain parameter $k$.

In fact, exploring the $n$ and $k$ parameter space of the core permittivity model shows that there is a region where the losses are overcome and a unique region where the super resonance, which is accompanied by a negative absorption cross-section, is achieved. Figure 13 shows contour and surface plots of the absorption cross-section, scattering cross-section and total cross-section over the $n$-$k$ parameter space for the resonance wavelength of 809nm. The super resonance is achieved only for values of $k$ corresponding to gain in the core, and for values of $n$ corresponding to the permittivity value of SiO$_2$: $\varepsilon_r = 2.05$, as well as the geometrical parameters needed for the resonances associated with the passive CNP.

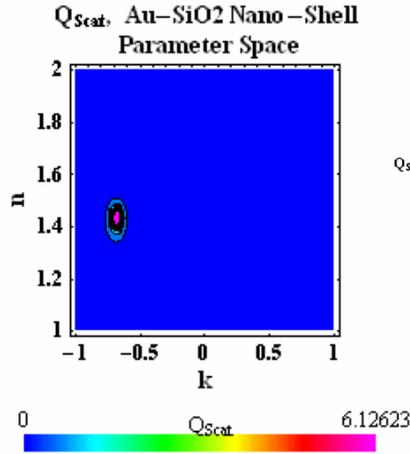
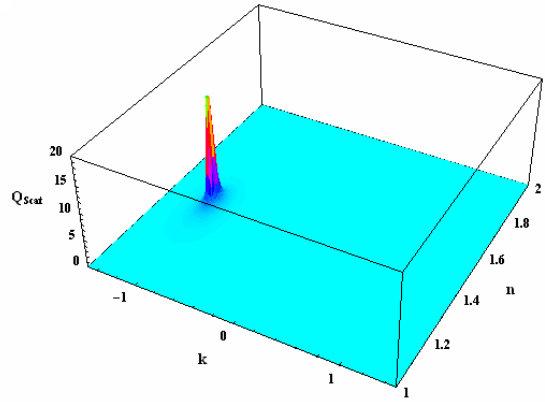
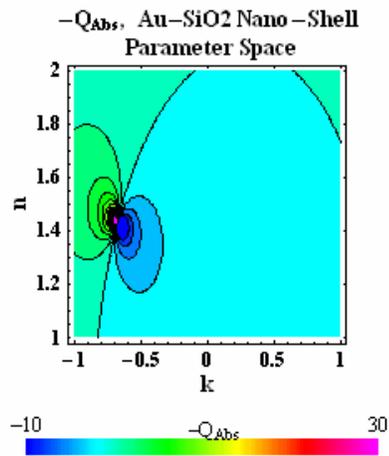
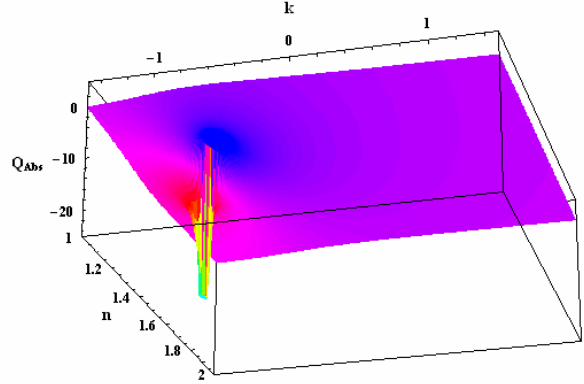

Fig. 13. Contour and surface plots of the scattering efficiency and the absorption efficiency as functions of *n* and *k*.

The extremely large increase in the radiated power and the drastic narrowing of the resonance peak at this super resonance suggests that the active CNP in this regime of the gain values has surpassed mere optical amplification of the incident beam and has become a CNP laser. However, from the exhibited behavior of the scattering and absorption efficiencies it is not clear whether the CNP is in fact acting like a laser resonator. In particular, is there energy being stored in the structure in the presence of the optical amplification? Thus, further investigations of the resonant characteristics in this super-resonance regime, such as the total amount of energy stored in the CNP, were made.

The total energy of the CNP was calculated as the sum total of the energy stored in shell and core regions. The electric and magnetic energy stored in each region was calculated with the relations:

$$W_E = \Re e\left\{ \oiiint_V \left[ \frac{1}{2}\partial_\omega(\omega\varepsilon)\left(|E_R|^2 + |E_\theta|^2 + |E_\phi|^2\right) \right] dV \right\} \quad (22)$$

$$W_H = \Re e\left\{ \oiiint_V \left[ \frac{1}{2}\partial_\omega(\omega\mu)\left(|H_R|^2 + |H_\theta|^2 + |H_\phi|^2\right) \right] dV \right\} \quad (23)$$

to account for the dispersive nature of the media involved in the CNP. In the optical regime, where there is a negligible magnetic material response, we have taken the permeability to be that of free space, $\mu = \mu_0$ in every region. Assuming the general case in which both the shell and core regions are modeled with complex dispersive permittivities and permeabilities, one finds that Eqs. (22) and (23) take the following forms in terms of the field coefficients in each region.

$$W_E^{Core} = \Re e\left\{ \frac{\partial_\omega(\omega\varepsilon_1)\pi r^2 |E_0|^2}{(\beta^2 - \beta^{*2})} \sum_{n=1}^{\infty} (2n+1) \left[ \begin{array}{l} A_1^{TECore}\Lambda_{jj(n)}(\beta r) \\ + B_1^{TMCore}\left\{\begin{array}{l}\left(\dfrac{n+1}{2n+1}\right)\Lambda_{jj(n-1)}(\beta r) \\ + \left(\dfrac{n}{2n+1}\right)\Lambda_{jj(n+1)}(\beta r)\end{array}\right\} \end{array} \right]_{\substack{\beta=\beta_1 \\ r=r_1}} \right\} \quad (24)$$

$$W_M^{Core} = \Re e\left\{ \partial_\omega(\omega\mu_1)\frac{|\sqrt{\varepsilon_1}|^2}{|\sqrt{\mu_1}|^2}\frac{\pi r^2 |E_0|^2}{(\beta^2 - \beta^{*2})} \sum_{n=1}^{\infty} (2n+1) \left[ \begin{array}{l} B_1^{TMCore}\Lambda_{jj(n)}(\beta r) \\ + A_1^{TECore}\left\{\begin{array}{l}\left(\dfrac{n+1}{2n+1}\right)\Lambda_{jj(n-1)}(\beta r) \\ + \left(\dfrac{n}{2n+1}\right)\Lambda_{jj(n+1)}(\beta r)\end{array}\right\} \end{array} \right]_{\substack{\beta=\beta_1 \\ r=r_1}} \right\} \quad (25)$$

$$W_E^{Shell} = \Re e \left\{ \frac{\partial_\omega(\omega\varepsilon_2)\pi r^2 |E_0|^2}{(\beta^2 - \beta^{*2})} \sum_{n=1}^{\infty} (2n+1) \begin{bmatrix} A_1^{TEShell}\Lambda_{jj(n)}(\beta r) + A_2^{TEShell}\Lambda_{yy(n)}(\beta r) \\ +B_1^{TMShell}\left\{ \begin{array}{l} \left(\dfrac{n+1}{2n+1}\right)\Lambda_{jj(n-1)}(\beta r) \\ +\left(\dfrac{n}{2n+1}\right)\Lambda_{jj(n+1)}(\beta r) \end{array} \right\} \\ +B_2^{TMShell}\left\{ \begin{array}{l} \left(\dfrac{n+1}{2n+1}\right)\Lambda_{yy(n-1)}(\beta r) \\ +\left(\dfrac{n}{2n+1}\right)\Lambda_{yy(n+1)}(\beta r) \end{array} \right\} \\ +2\,\mathrm{Im}\left\{ \begin{array}{l} A_3^{TEShell}\Lambda_{jy(n)}(\beta r) + \\ B_3^{TMShell}\left\{ \begin{array}{l} \left(\dfrac{n+1}{2n+1}\right)\Lambda_{jy(n-1)}(\beta r) \\ +\left(\dfrac{n}{2n+1}\right)\Lambda_{jy(n+1)}(\beta r) \end{array} \right\} \end{array} \right\} \end{bmatrix}_{\substack{\beta=\beta_2 \\ r=r_1}}^{\substack{\beta=\beta_2 \\ r=r_2}} \right\} \quad (26)$$

$$W_M^{Shell} = \Re e \left\{ \partial_\omega(\omega\mu_2) \frac{|\sqrt{\varepsilon_2}|^2}{|\sqrt{\mu_2}|^2} \frac{\pi r^2 |E_0|^2}{(\beta^2 - \beta^{*2})} \sum_{n=1}^{\infty} (2n+1) \begin{bmatrix} A_1^{TEShell}\left\{ \begin{array}{l} \left(\dfrac{n+1}{2n+1}\right)\Lambda_{jj(n-1)}(\beta r) \\ +\left(\dfrac{n}{2n+1}\right)\Lambda_{jj(n+1)}(\beta r) \end{array} \right\} \\ +A_2^{TEShell}\left\{ \begin{array}{l} \left(\dfrac{n+1}{2n+1}\right)\Lambda_{yy(n-1)}(\beta r) \\ +\left(\dfrac{n}{2n+1}\right)\Lambda_{yy(n+1)}(\beta r) \end{array} \right\} \\ +B_1^{TMShell}\Lambda_{jj(n)}(\beta r) + B_2^{TMShell}\Lambda_{yy(n)}(\beta r) \\ +2\,\mathrm{Im}\left\{ \begin{array}{l} A_3^{TEShell}\left\{ \begin{array}{l} \left(\dfrac{n+1}{2n+1}\right)\Lambda_{jy(n-1)}(\beta r) \\ +\left(\dfrac{n}{2n+1}\right)\Lambda_{jy(n+1)}(\beta r) \end{array} \right\} \\ +B_3^{TMShell}\Lambda_{jy(n)}(\beta r) \end{array} \right\} \end{bmatrix}_{\substack{\beta=\beta_2 \\ r=r_1}}^{\substack{\beta=\beta_2 \\ r=r_2}} \right\}$$

(27)

where the terms

$$\Lambda_{uv(n)}(\beta r) = (\beta * u_n v*_{n-1} - \beta u_{n-1} v*_n)$$
$$\Lambda_{uv(n+1)}(\beta r) = (\beta * u_{n+1} v*_n - \beta u_n v*_{n+1}) \quad (28)$$
$$\Lambda_{uv(n-1)}(\beta r) = (\beta * u_{n-1} v*_{n-2} - \beta u_{n-2} v*_{n-1})$$

The functions $u$ and $v$ in these terms are set to one of the spherical Bessel functions of the first or second kind, $j_n(\beta r), y_n(\beta r)$, as dictated by the subscripts on the factors $\Lambda_{uv(n)}(\beta r)$. The coefficients $A_1^{TECore}, B_1^{TMCore}, A_1^{TEShell}, B_1^{TMShell}, A_2^{TEShell}, B_2^{TMShell}, A_3^{TEShell}, B_3^{TMShell}$ are defined through the field expansion coefficients in each region, which can be determined by solving the matrix equation (9) for the CNP system; and $\varepsilon_i$ and $\mu_i$, with $i=1,2$, are, respectively, the permittivity and permeability in the core and shell region and $\beta_i = \beta_o \sqrt{\varepsilon_i/\varepsilon_0}\sqrt{\mu_i/\mu_0}$ is the propagation constant in each region.

The CNPs were designed to have their dipole scattering terms be resonant; thus keeping only the first four terms in the field expansions when evaluating expansion-based expressions such as Eqs. (24)-(27) is sufficient for accurate results. The following plot of the energy stored in the Au-SiO$_2$ CNP was obtained at the resonance wavelength of 809 nm.

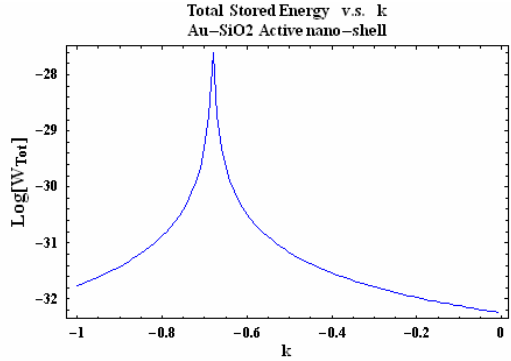

Fig. 14. Energy stored in the active Au-SiO$_2$ CNP as a function of the gain parameter *k*.

From these calculations it is clear that in comparison to the gain parameter values outside of the super resonance region, the stored energy within the CNP increases several orders of magnitude as the super resonance region is achieved. Moreover, one finds that large amounts of energy are contained within the active core of the CNP as compared to the plasmonic shell region. This can also be visualized with electric and magnetic field plots showing the field in the core, shell, and surrounding regions of the CNP at and away from the super resonance region. Electric and magnetic field plots for *k* values near to those that yield the super resonance are shown in Figs. 15 and 16 where the total field is plotted. Contour plots of the $E_\theta$ and $H_\phi$ components of the super resonance field are given in Fig. 16; they clearly demonstrate that the super resonance field is dominated by the dipole contributions, and that the fields within the plasmonic shell are considerably lower than in the core and near the outer surface of the CNP.

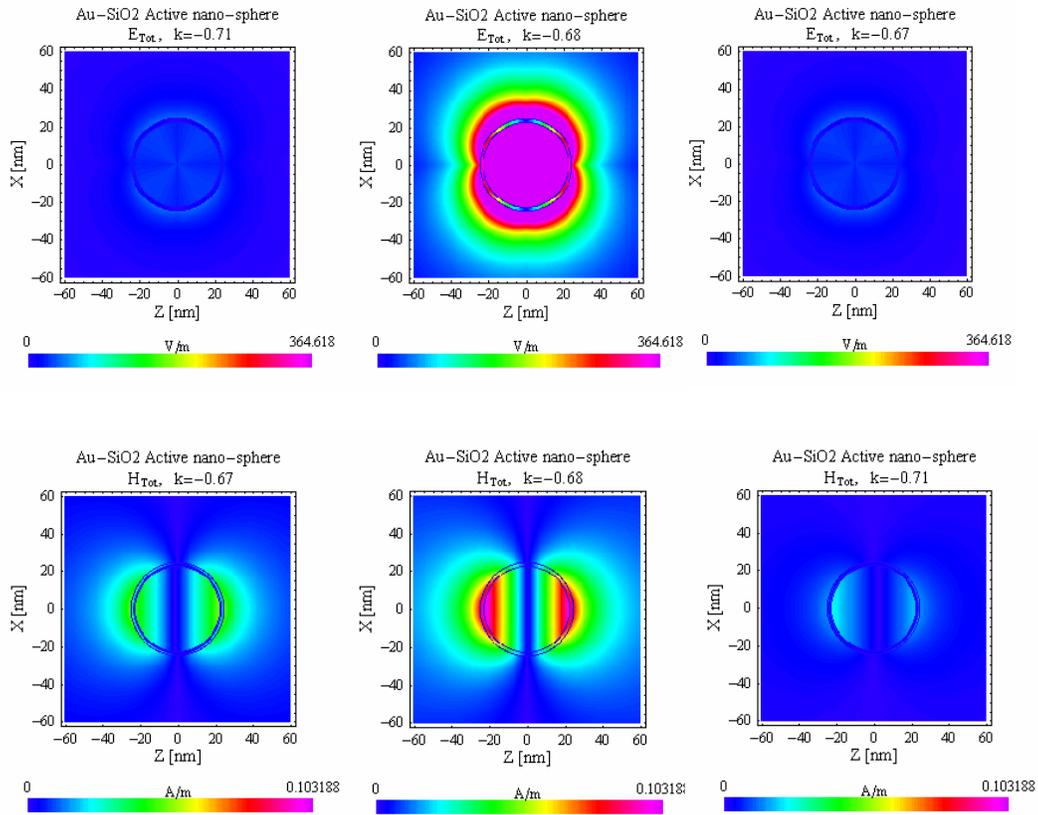

Fig. 15. The *total* electric and magnetic field distributions in the near-field region of the CNP with an active-$SiO_2$ core and an Au nano-shell for several values of the optical gain parameter *k*.

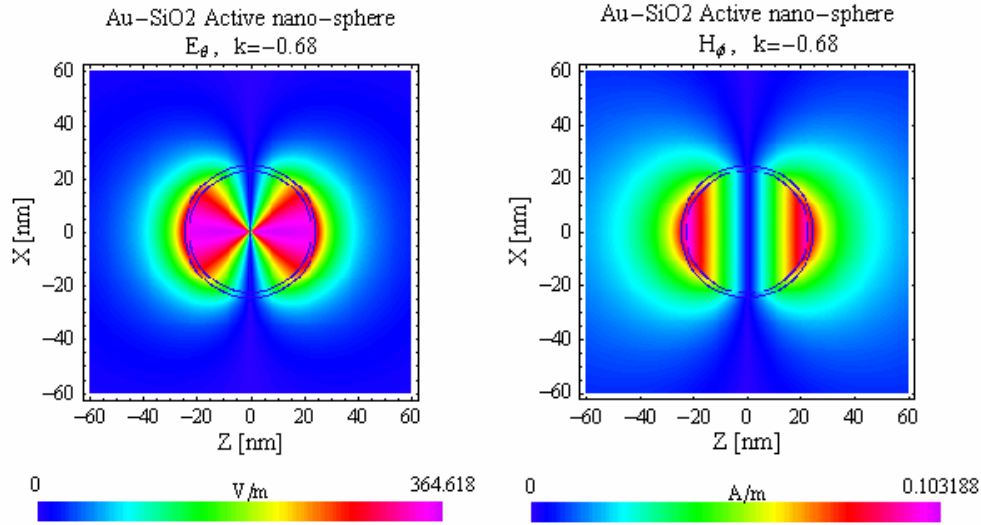

Fig. 16. Plots of the near-field distribution of the field components $E_\theta$ and $H_\phi$ for the CNP with an active-$SiO_2$ core and an Au nano-shell show that the dipole contributions dominate their behavior.

Similar to the scattering and absorption efficiency parameter space plots in Fig. 13, the total stored energy in the Au-$SiO_2$ CNP is shown as a function of $n$ and $k$ in Fig. 17. We observe, as we did in the efficiency plots, that there is a unique region where the stored energy is a maximum and that this maximum is several orders greater than the stored energy values determined in other regions of the parameter space. This maximum-stored-energy region coincides with those of the extrema in the scattering and absorption efficiencies. Consequently, the active CNP is lasing in this portion of the parameter space.

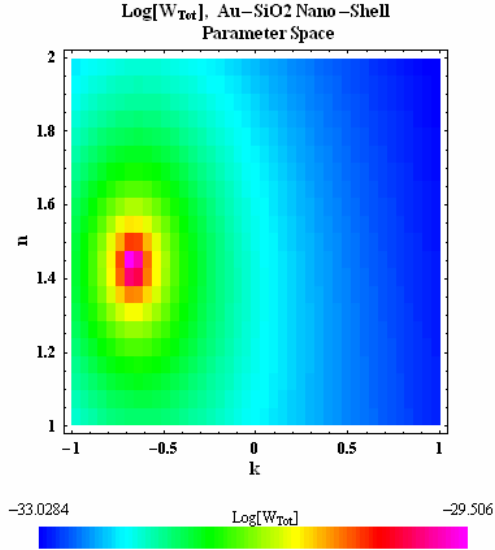

Fig. 17. A plot of the total stored energy in the Au-SiO$_2$ CNP as a function of *n* and *k*.

Next the results obtained using a rare-earth doped silica core, which emphasize the rare-earth ions being active at 510nm and represented by the susceptibility gain model in Eqs. (16)-(20), are presented. In exploring the use of such gain media, the products $N\sigma_{em}$ and $N\sigma_{abs}$, which represent the concentration of the rare-earth ions and their emission and absorption cross-sections, were varied in order to achieve sufficient gain to overcome the losses of the passive CNP. It was determined for the Ag-SiO$_2$ CNP that with an $N\sigma_{em} = 5.8 \times 10^4 \, cm^{-1}$ and $N\sigma_{abs} = 5.3 \times 10^4 \, cm^{-1}$, the needed pump power ratio value, *P*, for the population inversion of the doped silica was within reasonable limits, i.e., a few times the threshold value. The pump power was varied to achieve different gain values, the pump power ratio being constrained to the interval, $0 < P < 5$. The absorption and scattering cross-sections were then calculated as *P* was varied. The absorption and scattering efficiencies for gold and silver CNPs that have rare-earth-doped silica cores are shown, respectively, in Figs. 18 and 19.

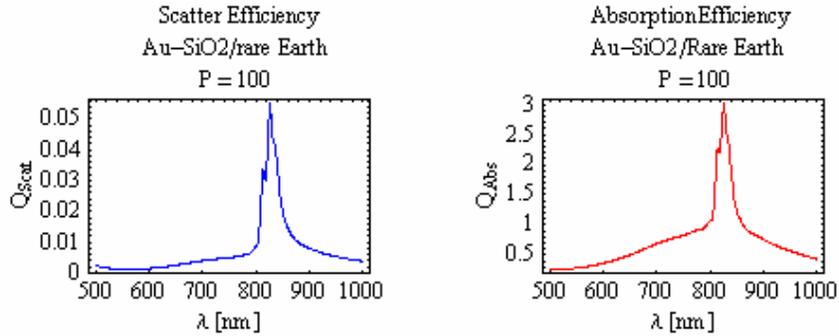

Fig. 18. Scattering and absorption efficiencies for the CNP with the Au shell and the rare-earth-SiO$_2$ core for the pump power ratio *P* = 100.

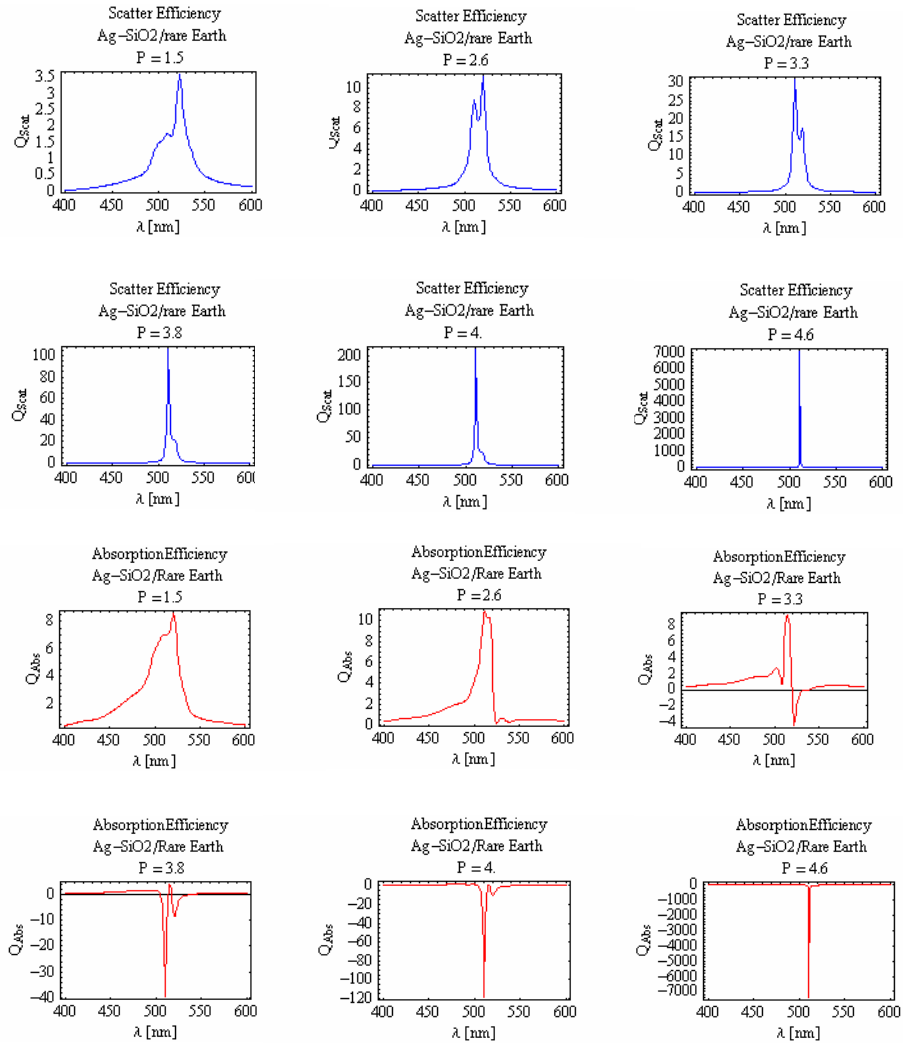

Fig. 19. Scattering and absorption efficiencies for the CNP with the Ag shell and the rare-earth-$SiO_2$ core for various values of the pump power ratio $P$.

From Fig. 18, one observes that the absorption efficiency remains positive and no significant loss compensation is achieved when the gold shell is used, even at $P$ values of 100. However, if a silver shell is used, one finds that the losses can be overcome for the same gain medium parameters, i.e., the lasing condition can be met, so that the super resonance is observed when $P = 4.6$. Figure 20 shows the net power leaving the active CNP at the peak of the resonance when the values of $P$ are varied. As was observed using the canonical gain model, it is clear that the rare-earth model shows a similar on/off feature of the lasing and the super resonance state.

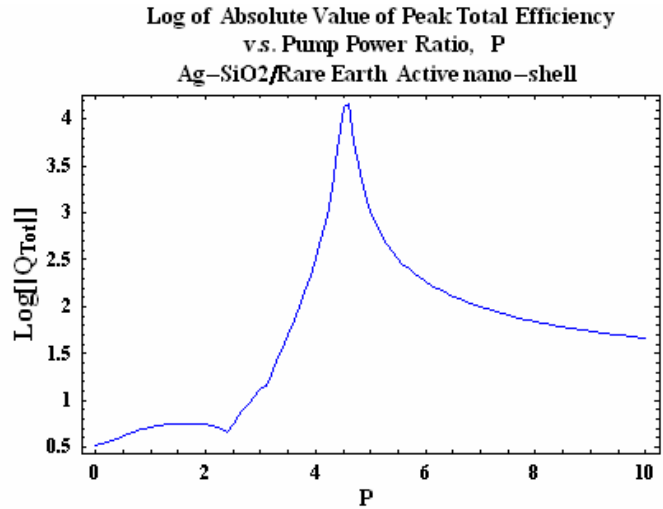

Fig. 20. The normalized total efficiency as a function of the pump power ratio $P$ for the Ag-SiO$_2$ CNP having a rare-earth core.

The stored energy plot for the rare-earth core CNP are shown in Fig. 21. As in the canonical gain core case, the extremum in the total stored energy as a function of the power parameter coincides with the maximally negative value of the absorption efficiency when the super resonance values is attained. Due to anomalous dispersion in the rare earth gain model used in the core region, the stored energy attains negative values passing through zero at the inflection point at $P = 1.6$. The energies are positive for $P > 1.6$.

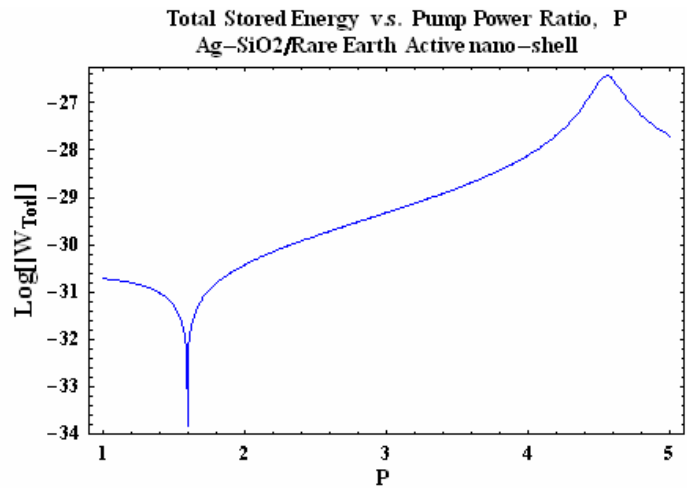

Fig. 21. The total stored energy in the Ag-SiO$_2$ CNP with the rare-earth core as a function of the pump power ratio.

Field plots for the rare-earth core CNP are shown in Figs. 22 and 23. Again the radiation is dominated by the dipole field and the highest fields are located at the surface of the CNP and within the core region with low field values occurring in the silver shell.

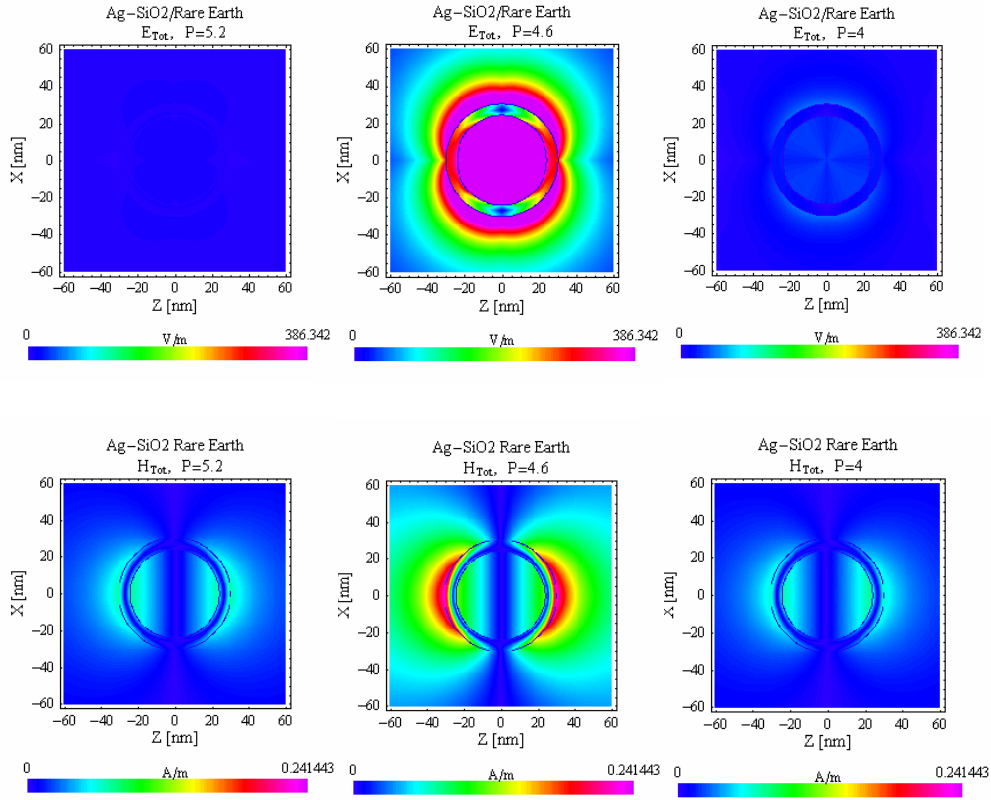

Fig. 22. The *total* electric and magnetic field distributions in the near-field region of the CNP with a rare-earth-SiO$_2$ core and an Ag nano-shell for several values of the pump power ratio *P*. Super resonance occurs at *P* = 4.6.

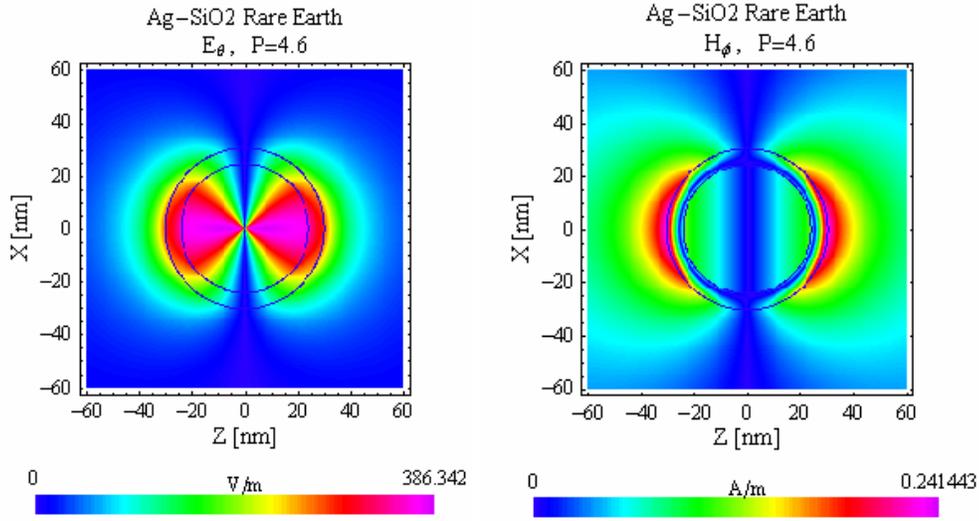

Fig. 23. Plots of the near-field distribution of the field components $E_\theta$ and $H_\phi$ for the CNP with a rare-earth-SiO$_2$ core and an Ag nano-shell show that the dipole contributions dominate their behavior.

## 5. Discussion

The parameter space plots for the canonical gain model suggest that when gain is added to the core of a plasmonic CNP, its losses can be overcome. In addition to the enhancements of the scattering and absorption cross-sections associated with the plasmon modes of the passive CNP, there are values of the permittivity of the core for which light amplification is possible. With doping densities on the order of $10^{20}$ cm$^{-3}$ and the $N\sigma \sim 10^4 \, cm^{-1}$ product used in our rare-earth gain models, the cross-sections that would be needed in practice to realize the reported CNP laser conditions are on the order of $10^{-16}$ cm$^2$. Recent studies with erbium-doped silicon nano-crystals [37] have achieved cross-section values of this order. In considering other forms of gain media that might be used in place of rare-earth ions, such as organic dyes or quantum dots, the above results obtained with the canonical gain model suggest that their optical gain coefficients must on the order of $\alpha \sim -10^4 \, cm^{-1}$. Furthermore, there are core permittivity values where phenomena indicative of lasing ensue. Within the quasi-static regime the sub-wavelength dimension of these nano-shells, $a \sim (25/500)\lambda = \lambda/20$, is well below the classical limiting dimension of $\lambda/2$ for a laser resonator. This behavior occurs because of the sub-wavelength resonances associated with an ENG coated sphere. Having the active ions in the core region appears to have several advantages over other configurations. We are currently investigating multi-layered nano-particles to determine whether there are other advantages or not to having multiple regions of passive and active media interacting with one another in an electrically small resonant nano-particle.

In classical laser operation the smallest cavity dimension defining the longitudinal laser modes is $\ell \sim \lambda/2$. As the pump power is increased past the threshold value needed for inverting the gain medium and overcoming the system losses in a classical laser below gain saturation, the laser output power increases monotonically. This behavior is due to the coherent oscillations of the resonant optical field in the laser cavity being amplified by the gain media. There is a one-to-one correspondence between the level of optical gain and the output optical power, but the feedback mechanism has no dependence on the optical gain inside the laser cavity. On the other hand, the coupled photon-plasmon polariton modes of the shell-core system in the CNP are in resonance. Therefore, if we consider the plasmon modes as providing a means for feedback for stimulated photons within the active core, the effects of the coherent oscillations of the electrons in the plasmonic shell on the core and shell permittivities must be considered. This implies that in an active CNP, an increase of the optical gain in the core does not necessarily lead to efficient coupling between the stimulated emission photons and the plasmon modes. Therefore as the gain is increased, the strength of the emission resonance may change. In particular, it may turn off the lasing state. To clarify this behavior, we considered a virtual mode analysis of the CNP.

In the presence of anharmonic material functions, the plasmon modes of the CNP system can be regarded as virtual modes with frequencies that in general take on values over the right half of the complex plane. With the interpretation that the imaginary frequencies associated with these virtual modes represent temporally growing or decaying states of the CNP, there is a connection between the material functions in the shell and core and the allowed virtual modes of the passive or active CNP. These virtual modes are excited in the absorption dominated passive CNP or the emission dominated active CNP. In isolating the mechanism responsible for the on/off nature of the super resonant lasing state in the CNP, we have begun investigating the influence of the material functions in the core and shell on the virtual modes of the CNP system in light of either growing or decaying modes of the active or passive CNP. The authors hope to explain the on/off nature of the super resonance lasing state of the active CNP from this perspective in a future publication.

## 6. Conclusions

In this paper we have presented the design and simulation of both passive and active plasmonic coated spherical nano-particles (CNP). The sizes of these CNPs were selected on the order of 20-30 nm to make them applicable to realizing optical metamaterials, as well as to investigate the possibility of realizing highly subwavelength resonant optical scatterers. The role of loss in the passive CNPs due to the optical properties of the plasmonic materials was considered, and the use of active materials in the design of CNPs was investigated to compensate this loss to achieve lossless active CNPs. In our simulations we have taken into consideration the size dependence of the plasmonic shells and have used both canonical gain models, as well as gain models representative of rare earth doped glass. From our investigations of active CNPs we have uncovered new phenomena that, aside from the geometric tunability of the passive CNPs, show a super resonant lasing state. In this super resonant state the CNP achieves a negative absorption efficiency of $10^3$ greater than the value obtained for a passive CNP, and a total efficiency of more than $10^4$. This increase in the total efficiency indicates the possible realization of a sub-wavelength laser whose size is on the order of $a \sim \lambda/20$. It was observed in the parameter space of the core and shell permittivities that there exists a well-defined region where the super resonance exists. Furthermore, the super resonant lasing state of the active CNP is observed to be localized in the gain parameter space in the sense that it can be turned on or off by adding gain values outside of the region in which the super resonance exists. This behavior must be contrasted with a classical laser where lasing is maintained as the gain is increased past its threshold value. It is believed that the lasing turns-off when the gain increases to such a point that it causes a severe detuning of the structural resonance. To explain the on/off nature of the super resonance, the authors are currently investing both lossy passive and active CNPs with a virtual mode analysis.

**Acknowledgments**

This work was supported in part by DARPA Contract number HR0011-05-C-0068.